\documentclass[10pt,journal,comsoc]{IEEEtran}
\ifCLASSOPTIONcompsoc
  \usepackage[nocompress]{cite}
  \usepackage{graphics, array, url, hyperref, epsfig, amsmath, amssymb, graphicx,color}
\usepackage{multirow}
\usepackage{subfigure}
\usepackage{threeparttable}
\usepackage{extarrows}
\else
  \usepackage{cite}
  \usepackage[bottom]{footmisc}
    \usepackage{graphics, array, url, hyperref, epsfig, amsmath, amssymb, graphicx,color}
\usepackage{multirow}
\usepackage{subfigure}
\usepackage{threeparttable}
\usepackage{extarrows}
\fi

\ifCLASSINFOpdf
\else
\fi
\begin{document}
%
\title{Security and Privacy for Mobile Edge Caching: Challenges and Solutions}
%
%
%

\author{Jianbing~Ni,~\IEEEmembership{Member,~IEEE,}
        Kuan Zhang,~\IEEEmembership{Member,~IEEE,}
       Athanasios V. Vasilakos,~\IEEEmembership{Senior Member,~IEEE}
\IEEEcompsocitemizethanks{
\IEEEcompsocthanksitem J. Ni is with the Department of Electrical and Computer Engineering, Queen's University, Kingston, Ontario, Canada K7L 3N6. Email: jianbing.ni@queensu.ca.\protect\\
\IEEEcompsocthanksitem K. Zhang is with Department of Electrical and Computer Engineering, University of Nebraska-Lincoln, Omaha, NE 68182 USA. Email: kuan.zhang@unl.edu.\protect
\IEEEcompsocthanksitem A. V. Vasilakos is with the School of Electrical and Data Engineering, University Technology Sydney, Australia, and with the Department of Computer Science, Electrical and Space Engineering, Lulea University of Technology, Lulea, 97187, Sweden. Email:th.vasilakos@gmail.com. \protect\\
}
}

%
%

\markboth{Journal of \LaTeX\ Class Files,~Vol.~14, No.~8, August~2015}%
{Shell \MakeLowercase{\textit{et al.}}: Bare Demo of IEEEtran.cls for Computer Society Journals}
%



\IEEEtitleabstractindextext{%
\begin{abstract}
Mobile edge caching is a promising technology for the next-generation mobile networks to effectively offer service environment and cloud-storage capabilities at the edge of
networks. By exploiting the storage and computing resources at the network edge, mobile edge caching can significantly reduce service latency, decrease network load, and improve user experience. On the other hand, edge caching is subject to a number of threats regarding privacy violation and security breach. {In this article, we first introduce the architecture of mobile edge caching, and address the key problems regarding why, where, what, and how to cache. Then, we examine the potential cyber threats, including cache poisoning attacks, cache pollution attacks, cache side-channel attacks, and cache deception attacks, which result in huge concerns about privacy, security, and trust in content placement, content delivery, and content usage for mobile users, respectively. After that, we propose a service-oriented and location-based efficient key distribution protocol (SOLEK) as an example in response to efficient and secure content delivery in mobile edge caching. Finally, we discuss the potential techniques for privacy-preserving content placement, efficient and secure content delivery, and trustful content usage, which are expected to draw more attention and efforts into secure edge caching. }

\end{abstract}

}

\maketitle

\IEEEdisplaynontitleabstractindextext

%
\IEEEpeerreviewmaketitle

{\section{Introduction}\label{sec:introduction}}
Due to the vastly increased number of connected devices in the Internet of Things (IoT) \cite{Stegiou18}, wireless networks are expected to provide supremely high data rates and extremely low latency to enable the emerging applications, such as remote control of smart vehicles, industrial automation, and virtual/augmented reality. By integrating the disruptive technologies, mobile devices have ultra-reliable and affordable network access to satisfy the quality of experience in wireless environments, and thereby making our life more convenient. As one of the key technologies, multi-access edge computing \cite{Hu15} is becoming increasingly popular that offers service environment and cloud-computing capabilities in an effective manner at the edge of wireless networks. By leveraging the computing and storage resources on the edge nodes (e.g., macro base stations, Wi-Fi access points, network switches, video cameras, and roadside infrastructure), popular contents can be temporarily maintained on the
edge nodes to respond to frequent access requests from user equipment, which is referred to as
mobile edge caching \cite{Zeydan16}. Since the contents are delivered directly from edge nodes instead of from the remote cloud if they are cached on edge nodes when requested, network edge caching can significantly reduce the content delivery
latency, alleviate the backhaul traffic, and improve the spectrum efficiency and the energy efficiency
\cite{Liu16}.

Despite the appealing benefits of the quality of service, caching at the network edge has been subject to a number of controversies regarding security breach and privacy violation. For example, Denial-of-Service (DoS) attacks and wireless jamming can consume the bandwidth and computing resources at the network edge. The open edge network creates the real-world attack surface for adversaries that aim to compromise or control the edge caches. The attackers might acquire the stored data at the edge caches that are located at their vicinity, as well as the data that are from other locations in distributed edge storage, such that the sensitive information can be extracted because of the context awareness. In addition, the edge network providers might be curious about the contents stored on their caches and utilize the inference attacks to identify the hidden information about mobile users, such as trajectory and personal identifiable information. However, with limited computation, power, and memory, complex secure and privacy-preserving mechanisms cannot be straightforwardly deployed on the edge caches. Therefore, it is of paramount importance to explore deeper understanding of the potential attacks and design practical real-world solutions for reasoning about the security and privacy of mobile edge caching.

{In this article, we aim to examine security and privacy threats in mobile edge caching, and explore new models, techniques, and approaches to reason about the security, privacy, and trust of mobile edge caching. Specifically, we start with the architecture of mobile edge caching and address the key issues about edge caching, i.e., why to cache, where to cache, what to cache, and how to cache. We then examine a series of threats on security and privacy that are grouped into two categories: 1) generic threats, such as DoS attacks, wireless jamming, malware attacks, and man-in-the-middle attacks; and 2) cache-specific threats, including cache poisoning attacks, cache pollution attacks, cache side-channel attacks, and cache deception attacks. However, the research about edge caching security is still in the early stage. We discuss the privacy, security, and trust challenges in mobile edge caching, and explore the potential solutions to preserve privacy in content placement, security in content delivery, and trust in content usage. To identify the approaches of securing content delivery, we propose a \underline{S}ervice-\underline{O}riented and \underline{L}ocation-based \underline{E}fficient \underline{K}ey distribution protocol (SOLEK) for secure content sharing based on the certificateless proxy re-encryption as an example. Finally, we discuss future research directions of privacy-preserving content placement, efficient and secure content delivery, and trustful content usage, to draw more efforts for securing edge caching.}

\section{Overview of Network Edge Caching} \label{sec2}
In this section, we present the general architecture of mobile edge caching. As shown in Fig. 1, mobile edge caching involves several entities, i.e., user equipment, access networks, mobile core network, and the Internet. User equipment (UE), such as smart phones, smart vehicles, smart meters, and smart home produces, carried by mobile users or deployed in ground, are capable of accessing different services in the Internet through access networks and mobile core networks. In access networks, different wireless communication technologies, including radio access network, Wi-Fi, small cells, and 5G new radio, are leveraged to build communication channels for UE with the support of infrastructure, such as base stations, edge routers, and switches. Mobile core network is the heart and acts as an anchor point for multi-access technologies. 5G core comprises of pure, virtualized, software-based network functions or services, and can be instantiated within multi-access edge-cloud computing infrastructures. The Internet is the data network that provides a large variety of services to the connecting mobile users.

\begin{figure}
\centering
\includegraphics[width=0.5\textwidth]{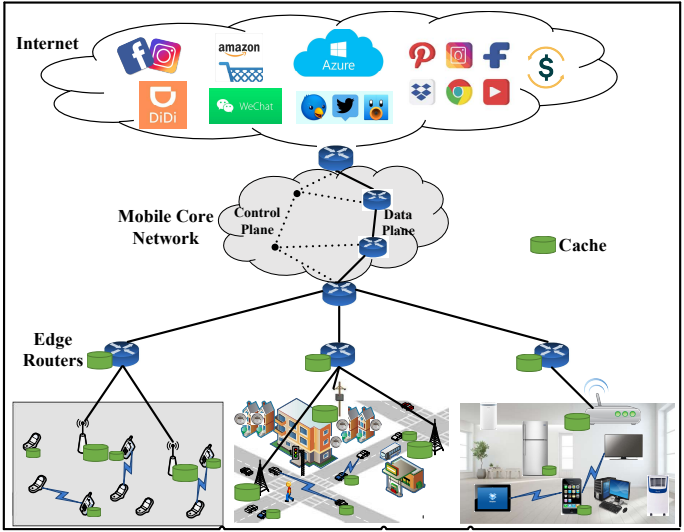}
\caption{Mobile Edge Caching.}
\label{fig:one}
\end{figure}

The problems of why, where, what, and how to cache are the key issues in mobile edge caching. Why to cache refers to the advantages of mobile edge caching regarding the quality of experience of mobile users. Where to cache refers to location selection for caching contents. What to cache refers to the caching contents that have short lifetimes and consists of different types, such as documents, videos, radios, and IoT data, and how to cache refers to caching policy selection and caching scheme design.

\textbf{Why to cache?} The primary reason for caching is to make data access less expensive on time, bandwidth, and resource costs.
\begin{itemize}
  \item Service Delay Reduction: By storing content close to mobile users, the service delay can be significantly reduced. This feature is quite attractive for data delivery of latency-sensitive applications, such as video streaming that will account for 78\% of all mobile traffic by 2021.
  \item Traffic Burden Releasing: As duplicate contents cause severe traffic burden over hackhaul links, the traffic can be reduced up to 35\% by storing the requested contents at the network edge \cite{Hu15}. Caching in 3G and 4G LTE networks can reduce mobile traffic up to two thirds \cite{Wang14}.
  \item Energy and Spectral Efficiency Improvement: Energy and spectral efficiency can be improved by caching contents at user devices, or by caching at base stations or edge routers to eliminate the backhaul bottleneck. Spectral efficiency and energy efficiency can benefit from wireless edge about 900\% and 500\%, respectively \cite{Liu16}.
\end{itemize}

\textbf{Where to cache?} As depicted in Fig. 1, the caches can be placed on the devices at the network edge, including edge in wired network and edge in wireless network.

\begin{itemize}
\item Caching at UE: Caching at UE is referred to device-to-device caching. The devices are organized into clusters and directly communicate with each other using license-band or unlicense-band protocols. The requested content can be offered by some other UE within the same cluster that are caching the content.

\item Caching at Base Stations (BSs): Caching at BSs is referred to deploy caching at micro base stations (MBSs), small base stations (SBSs), pico base stations (PBS), and femto base stations (FBSs). Generally, MBSs have larger cache spaces and coverage areas than SBSs, and SBSs use cooperative caching to share contents with each other. PBSs and FBSs are special SBSs. In 5G era, C-RAN is an important architecture of 5G cellular network \cite{Yang17}. C-RAN aggregates the computing capabilities to baseband unit (BBU) pool, and deploys multiple radio remote heads (RRHs) to address network capacity and coverage issues. Cache spaces are deployed at both BBU and RRH that form a distributed edge caching architecture. Besides, at wireless network edge, caching can be located at wireless routers for enabling various applications, e.g., smart home and smart office.
\end{itemize}

In addition, the contents can be cached at wired network edge, i.e., on edge routers.

\textbf{What to cache?} The aim of caching is to reduce backhaul traffic and service delay, so that the network providers prefer to cache all the contents that may be requested by users. Nevertheless, the storage spaces of edge caches are limited and the scale of contents is growing significantly \cite{Liu16}. Thus, it is important to determine what content to be cached by considering popularity. In fact, only a small number of content is extremely popular, while a long tail of content is accessed by a small portion of mobile users. ``Cacheability" is changing over caching strategies and content types \cite{Wang14}. For example, it is unnecessary to cache the contents that have been maintained on the neighboring caches, and some contents with special types, e.g., images and videos, may have the highest priority to be cached as they have the highest revisit rate. Moreover, only public and static content, such as images, CSS files, PDF documents, and videos will be cached. The cached contents are not user-specific.

\textbf{How to cache?} {According to when to decide whether to cache contents on edge nodes, caching can be classified as reactive caching and proactive caching. Reactive caching determines whether to cache a particular content after it has been requested, while proactive caching determines what contents to be cached before they are requested. Both reactive caching and proactive caching make the decision based on the content types (e.g., files, software packages, and videos), content popularity (i.e., the ratio of the number of requests in a particular region), or the profiles of potential interested users (e.g., access patterns and mobility patterns). The content popularity and the profiles of interested users may vary with time. The cached contents would be updated based on the caching policies or strategies. In general, the proactive caching and reactive caching include the following steps:}

\begin{itemize}
  \item {Content Placement/Update: The objectives of content placement and content update are consistent, i.e., to determine locations of contents for maximizing cache-hit rate. An effective caching policy can accurately estimate the gains for caching content by evaluating current and future popularity and optimizing storage spaces and cache node selection. Therefore, in content placement and content update, one of the key issues is to estimate the popularity of content. The user mobility, social associations, and preferences are essential criterions to predict content's popularity, and enable effective content placement and cache update to achieve high cache-hit rate.}
  \item {Content Delivery: The UE requests the interested contents from the service servers. After receiving content requests, edge nodes determine whether the requested content is cached or not by finding the unique identifier for every object, called cache key, in the cache. If a cache-hit happens, the edge node directly delivers the requested content to UE; otherwise, it fetches the content from the remote servers through the mobile core network.}
  \item {Content Usage: The UE receives the requested contents from the edge nodes or the remote servers. The UE needs to check that the contents are latest and they are from the correct service providers. The UE not only makes use of the obtained contents to enjoy the corresponding services, but also serves as the cache node to offer contents to other UE within the cluster.}
\end{itemize}

\section{Privacy, Security, and Trust} \label{sec3}
In this section, we introduce the potential cyber threats that corrupt security, privacy, and trust in mobile edge caching, and discuss the issues caused by these threats.

\subsection{Threats in Mobile Edge Caching}
With the development of ubiquitous networking, the connected devices are exposed to various cyber threats, ranging from generic threats, such as DoS attacks, wireless jamming, malware attacks, and man-in-the-middle attacks, to cache-specific threats, including cache poisoning attacks, cache pollution attacks, cache side-channel attacks, and cache deception attacks. The generic threats to edge networks are the same as the risks to the well-known communication and information systems. For example, the DoS attacks flood edge nodes with superfluous requests in an attempt to make the services offered by the edge nodes unavailable to its intended UE. We omit the introduction of generic attacks and refer to \cite{Lu20} for the details. The cache-specific attacks are introduced as follows.

\textbf{Cache poisoning attacks}: {An adversary inserts the forged or false contents into the edge caches to deceive the users who request them. For example, by utilizing the vulnerability of cache keys, an adversary creates a cache key collision and manipulates the cache keys to cause a harmful or crafted response to the request from the targeted UE. The attacker may request a content with a query containing a malformed identifier, which may cause the remote server get crash. As a consequence, the false content is returned and cached on the edge node and served other mobile users who request it. The cache poisoning attack could result in an attacker delivering false or error HTML pages, javascript files, stylesheets, images, and videos to the mobile users instead of legitimate contents. Therefore, the cache poisoning attack threatens the availability and the correctness of network resources.
The users may be successively deceived until the poisoned cache is updated. The success of this attack relies on the exploitable vulnerabilities on the edge caches.}

\textbf{Cache pollution attacks}: An attacker pollutes caching balance by sending large numbers of fabricated contents to the cache and thereby tricking the cache into caching non-popular contents. Under this attack, the cache-hit rate of the legitimate users is degraded, and their response time is increased. There are two types of cache pollution attacks, i.e., location-disruption attacks and false-locality attacks. In the location-disruption attack, an adversary frequently requests some non-popular contents and squeezes out the popular contents in caches. In the false-locality attack, an adversary continually requests a set of non-popular contents and inserts them into the cache.

\textbf{Cache side-channel attacks}: By observing and measuring the activities of edge caches, such as response time, power consumption, cache access, and returned fault, an adversary can learn private information about mobile users and cached contents. The cache side-channel attacks can be divided into different types based on the knowledge that the adversaries have. For example, in cache-timing attacks, by comparing the response time, an adversary can determine whether particular content has been cached. In content deduplication attacks, a copy-on-write content fault reveals the fact that the requested content is deduplicated and users must have requested the identical content. In cache privacy attacks, an adversary can learn the access history and other possible private information of a particular user, if they share the same edge cache.

\textbf{Cache deception attacks}: The attacker firstly tricks the target user into making a request to the private content that it is not permitted to access. The attacker then submits the same request, and edge caches can return the requested content containing the previously private data. For example, an attacker crafts a URL request that points to the bank account information of a victim, but appends to it a non-exist path of a static image, and tricks the victim into making this request. The server will ignore the invalid suffix and return the account details, and the cache node will maintain the content which is identified as a static image. The attacker can make the same request to retrieve the account details. Thus, the private data of the victim will be disclosed.

{The cache poisoning attacks and the cache pollution attacks may corrupt the integrity of cached content, render the unavailability of requested data, and degrade the trust of mobile users on the edge caching services. The cache side-channel attacks and the cache deception attacks compromise the confidentiality of contents on caches and invade the privacy of mobile users. These threats on security, privacy, and trust may compromise all the appealing benefits of mobile edge caching. Security, privacy, and trust are different requirements for the reliability of mobile edge caching, but they are all interrelated. If security or privacy are not protected in content placement or content delivery, the trust in data usage cannot be built. Therefore, it is essential to preserve security, privacy, and trust in the entire procedure of mobile edge caching services.}

\subsection{Security, Privacy, and Trust Challenges}
We examine the typical issues in the whole procedure, from content placement to content delivery and content usage.

\textbf{Privacy in Content Placement}: Compared to the increasing data volumes in mobile networks, the storage space on edge nodes is always limited. It is impossible to cache all data locally. Hence, identifying the optimal set of contents that maximizes the utility of each cache becomes crucial. A common approach of content set identification for placing contents is based on content popularity. To estimate popularity, a machine-learning model \cite{Xiao18,Wang19} is needed to analyze historical access records, social association, and mobility patterns of mobile users. However, machine learning is a double-edged sword. While enabling the discovery of knowledge from a large volume of data, it may invade user privacy by exposing potentially sensitive information. Such privacy invasion may trigger serious results. For example, the exposure of mobile users' preferences may bring various unwelcome advertisements and promotions. Therefore, questions that need to be addressed include, how to achieve privacy-preserving content placement based on content popularity and user preferences? How to predict content popularity using machine learning without exposing the privacy of mobile users? It is essential to explore privacy-preserving techniques and design advanced solutions as a long-term vision for content placement in mobile edge caching.

\textbf{Security in Content Delivery}: UE searches and retrieves contents from the correct caches if they are on edge nodes; otherwise, it has to request the data from remote servers. Knowing what content is stored on the edge nodes at any given moment is essential, while maintaining this knowledge is challenging in a distributed cache network. The Bloom filter and its variants are common approaches for space-efficient approximate membership representation, a useful technique to indicate whether an element is a member of a large group. A shortcoming of such approaches is the false positive rate, i.e., a given item is indicated to be cached but actually not there. Othello hashing is used to design a memory-efficient and fast localization method of edge nodes \cite{Cohen19} to reduce the rate of false locating, but Othello construction and update operations are expensive. Besides, end-to-end encryption is the major method of securing data transmission on public channels (e.g., the transport layer security (TLS) protocol), while protecting contents maintained on edge nodes against the cache side-channel attacks and the cache deception attacks, but it seemingly signifies the death of edge caching. The contents cached on edge nodes are encrypted for the first requesting mobile user, such that the subsequent UE cannot decrypt the fetched data from the edge nodes. CryptoCache \cite{Leguay17}, a secure content transmission protocol was designed based on symmetric encryption to enable caching of encrypted contents on an untrusted edge node, but the cloud server is still needed even if the requested contents are cached. Content partitioning and scrambling \cite{Su18} is an efficient solution to achieve content confidentiality but cannot provide sufficient security guarantees, i.e., semantic security; and attribute-based encryption \cite{Pan19} offers high data security, but brings heavy computational overhead. Therefore, questions include, how to design effective content locators on edge nodes for optimal content exploration? How to design efficient data encryption schemes for secure content delivery without involving remote cloud servers? Exploring new models, techniques, and approaches to guarantee effective content exploration and reason about the data confidentiality during content delivery against the cache side-channel attacks and the cache deception attacks is deserved to focus on.

\textbf{Trust in Content Usage}: Mobile users recover their contents of interest after receiving the encrypted data from edge caches. {Due to the security threats, mobile users may have huge concerns regarding the trustworthiness of their obtained contents, e.g., whether they have been corrupted by ill-intentioned edge nodes, poisoned by adversaries with cache poisoning attacks, or are unavailable when requested due to the cache pollution attacks. The trust decline due to the violation of authenticity, integrity, and availability is one of the major obstacles for the widespread of mobile edge caching.
The sophisticated techniques (e.g., content checking, message authentication, and digital signature) may not be practically feasible to address these issues as a whole. For example, the digital signature can guarantee the authenticity and integrity of the content, but it cannot effectively prevent or detect cache poisoning attacks, since in the cache poisoning attack, the response is from the correct server, but the content is not the one requested.}
Moreover, they bring extra computational and communication overheads to both edge nodes and mobile users. These additional burdens may eventually ruin the major advantage of edge caching, i.e., low-latency service response. Without efficient integrity verification, the trustworthiness of cached contents depends on the honesty of edge nodes, but trust evaluation in a distributed edge network is not straightforward. Blockchain can potentially be leveraged to achieve distributed trust management on behalf of a decentralized, transparent and public ledger, but this research area remains a series of inherent issues, e.g., privacy leakage and storage overload. Potential questions include, how to design efficient content verification schemes to detect corrupted content, and how to build a trustful edge-caching framework based on the blockchain and smart contracts. These problems, if not properly resolved, will impede the success of mobile edge caching.


\section{SOLEK}  \label{sec5}
We propose SOLEK, a service-oriented and location-based efficient key distribution protocol based on certificateless proxy re-encryption \cite{Selvi17} for secure content delivery in mobile edge caching. SOLEK enables a mobile user to decrypt the ciphertexts of the contents cached on the edge node. The decryption capability is delegated based on the service type $ID_S$ and the position of the edge node $ID_i$, such as GPS data. {Specifically, a mobile user is distributed with two separate secret-public key pairs. One is related to the service type $ID_S$ that is given in the service registration, and the other is about the edge node $ID_i$ that is allocated when entering the coverage area of $ID_i$. Also, the mobile user can access the contents cached by a neighboring edge node $ID_j$ of $ID_i$ by utilizing the proxy re-encryption technique. The model of SOLEK is shown in Fig. 2, and the detailed algorithms in SOLEK are described in Fig. 3.}

\begin{figure}
\centering
\includegraphics[width=0.5\textwidth]{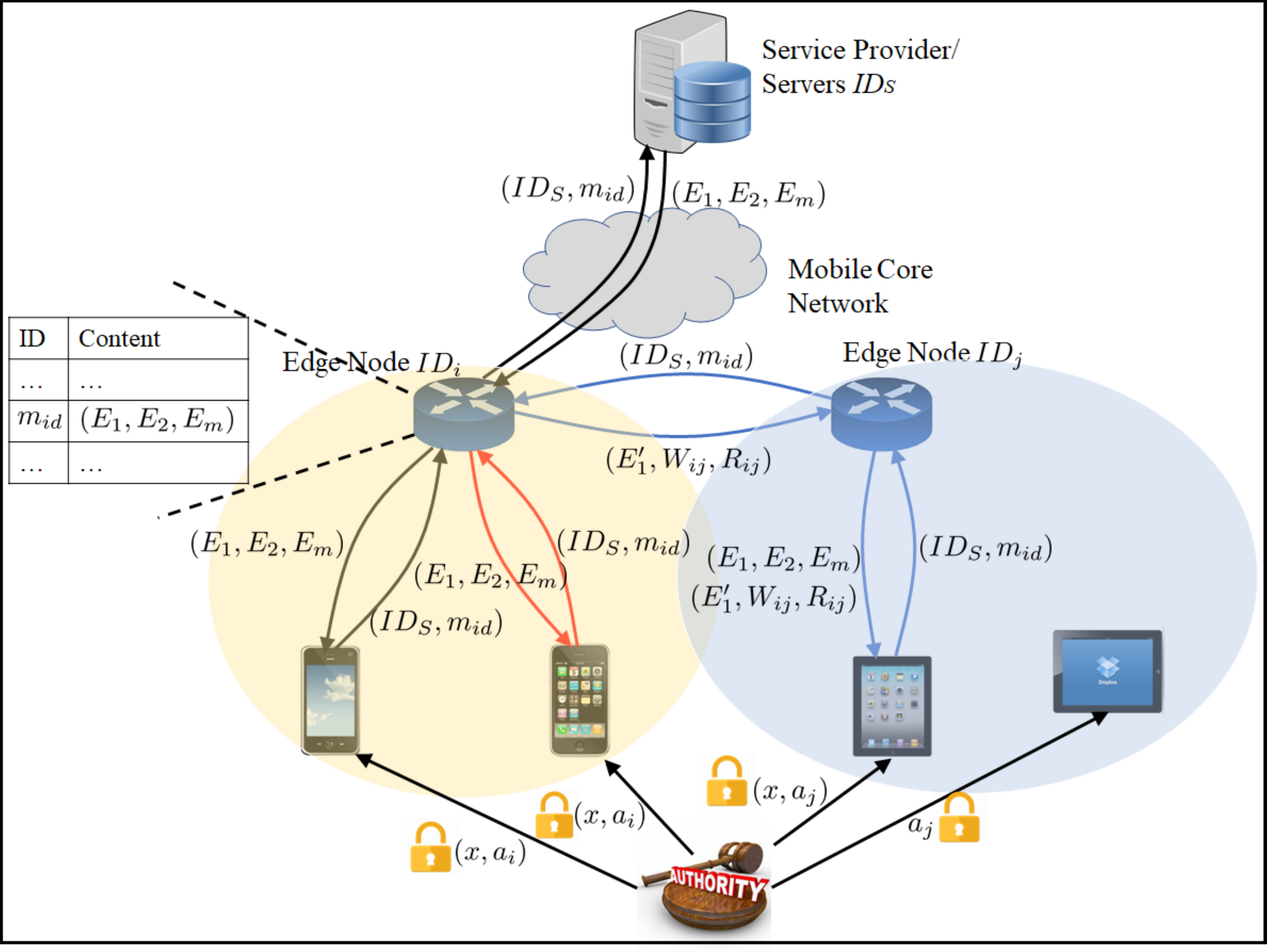}
\caption{Model of SOLEK}
\label{fig:one}
\end{figure}

\begin{figure*}
\centering
\includegraphics[width=1.0\textwidth]{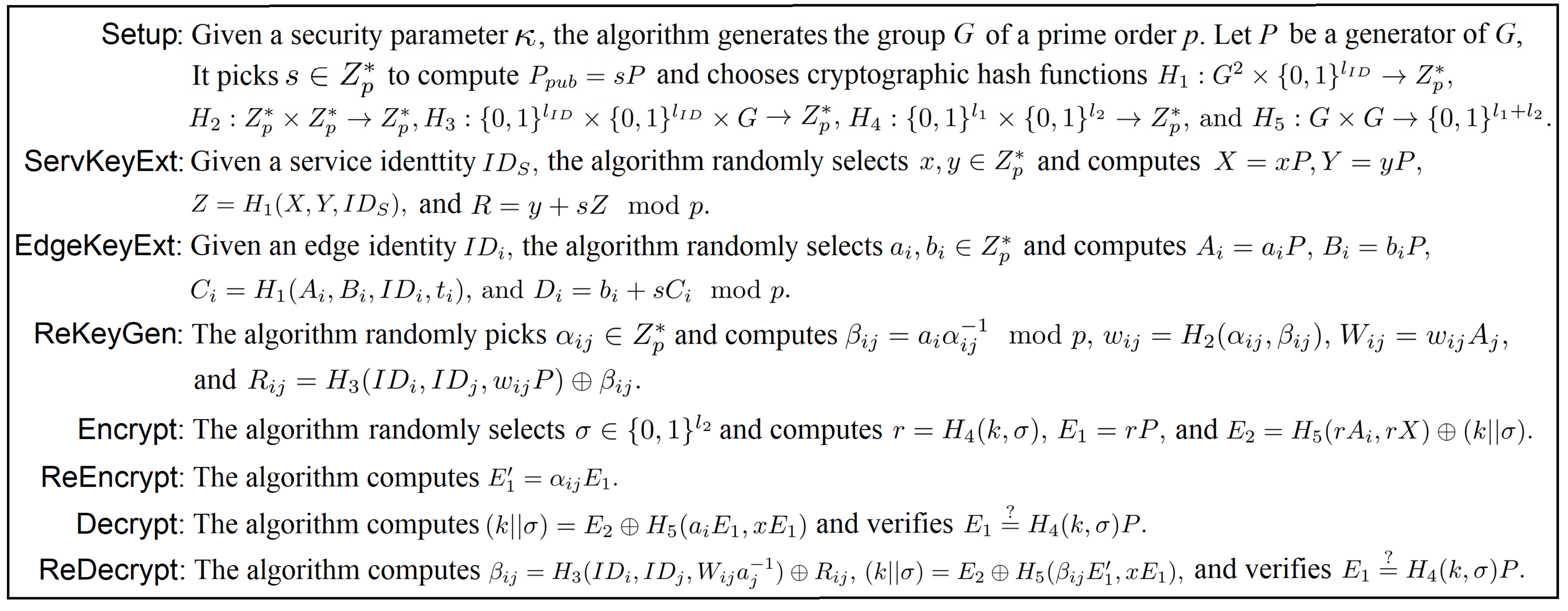}
\caption{SOLEK Algorithms}
\label{fig:one}
\end{figure*}

In SOLEK, a trusted authority (TA) first executes the {\sf Setup} algorithm to bootstrap the whole protocol. The TA can be the key generation center (KGC) in the identity-based cryptosystem. With the input of the security parameter $\kappa$, the {\sf Setup} algorithm outputs the public parameter $param=(G,p,P,P_{pub},H_1,H_2,H_3,H_4,H_5)$ and the master secret key $s$. Then, with the inputs $(param, s, ID_S)$, TA runs the {\sf ServKeyExt} algorithm to generate the service secret key $y$ and the service public key $SPK=(X,Y,R)$. $SPK$ is public and $y$ is sent to the mobile users who register the service $ID_S$ through secure channels. When receiving $y$, the mobile users can verify whether $y$ is correct by checking $RP \overset{?}{=} Y+H_1(X,Y,ID_S)P_{pub}$. Also, with the inputs $(param, s, ID_i)$, TA executes the {\sf EdgeKeyExt} algorithm to generate the edge secret key $a_i$ and the edge public key $EPK=(A_i,B_i, D_i)$ for the edge node $ID_i$. $EPK$ is public and $a_i$ is sent to the mobile user who enters the coverage area of $ID_i$ through a secure channel. When receiving $a_i$, the mobile user can verify whether $a_i$ is correct by checking $D_iP \overset{?}{=} B_i+H_1(A_i,B_i,ID_i,t_i)P_{pub}$. Here, $t_i$ is the current time slot, and the edge secret-public key pair can be updated by TA periodically. If a mobile user $U$ requests the data $m$ of the service $ID_S$ in the coverage area of $ID_i$, $U$ sends $ID_S$ and the file name $m_{id}$ to $ID_i$. $ID_i$ checks whether $m$ is cached or not. If not, i.e., cache miss, the request is forwarded to the server.
The server chooses a session key $k$ to encrypt $m$ to generate $E_m$ using a block cipher, e.g., AES, and runs the {\sf Encrypt} algorithm to encrypt $k$ to obtain $(E_1,E_2)$. $(E_1,E_2, E_m)$ is returned to $U$ and also cached on $ID_i$. $U$ runs the {\sf Decrypt} algorithm to decrypt $(E_1,E_2)$ and obtain $k$ for data decryption. If yes, it means cache hit, that is, $(E_1,E_2, E_m)$ is cached on $ID_i$, $U$ can directly fetch $(E_1,E_2, E_m)$ from $ID_i$. Subsequently, if $m$ is requested by anther mobile user $U'$ in the coverage area of $ID_j$, a neighboring edge node of $ID_i$, $ID_i$ runs the {\sf ReKeyGen} algorithm to generate the re-encryption key $RK=(\alpha_{ij},W_{ij},R_{ij})$, and the {\sf ReEncrypt} algorithm to transform $(E_1,E_2, E_m)$ to $(E_1,E'_1,E_2, W_{ij},R_{ij}, E_m)$ for $U'$. Finally, $U'$ fetches $(E_1,E'_1,E_2, W_{ij},R_{ij}, E_m)$ from $ID_i$, and runs the {\sf ReDecrypt} algorithm to decrypt $(E_1,E'_1, E_2)$ and obtain $k$ for data decryption.

{In SOLEK, two properties are provided, i.e., content confidentiality and content sharing.} To ensure the content confidentiality in SOLEK, two types of attackers should be considered. The one has registered the service $ID_S$ but is not in the coverage area of $ID_i$; the other is in the coverage area but does not register the service $ID_S$. If neither can compromise SOLEK, other attackers, e.g., non-registered $ID_S$ and outside $ID_i$ attackers, cannot break SOLEK. SOLEK has two levels of ciphertexts, $(E_1,E_2, E_m)$ and $(E_1,E'_1,E_2, W_{ij},R_{ij}, E_m)$. We ensure the confidentiality of cached content, i.e., each type of attackers cannot gain any knowledge from any level of ciphertexts. In order to formalize plaintext leakage from ciphertexts, the distinguishability has been defined that an attacker cannot identify the plaintext choice given an encryption of a plaintext randomly chosen from a two-element plaintext space determined by the attacker with probability significantly better than that of random guessing. According to this property, the security of SOLEK can be reduced to the hard assumption, i.e., the Decisional Diffie-Hellman (DDH) assumption.

SOLEK achieves efficient content sharing among mobile users in the same area for the service of common interest. It provides higher security guarantees than the scheme in \cite{Su18}, under the assumption that all the registered mobile users will not proactively expose their interested contents. Also, SOLEK avoids interactions with the server to retrieve the session key if cache hit compared with CryptoCache \cite{Leguay17}. Attribute-based encryption (ABE) provides tremendous potential for fine-grained data sharing based on multiple attributes, while SOLEK surpasses the ABE-based data sharing schemes, such as PCWX \cite{Pan19} and PWXL\cite{Pu19}, in terms of computational efficiency. The time costs of content encryption and decryption in PCWX \cite{Pan19}, PWXL\cite{Pu19}, and SOLEK are demonstrated in Fig. 4. Since the cached content is shared based on the service type and edge location, two attributes are considered in PCWX \cite{Pan19} and PWXL\cite{Pu19}. The experiments are conducted on a smartphone of HUAWEI MT2-L01 with Kirin 910 CPU and 1.25MB memory. The Barreto-Naehrig curve is utilized and the security parameter $\kappa=128$. {Fig. 4(a) and Fig 4(b) demonstrate that SOLEK is much more efficient than PCWX \cite{Pan19} and PWXL\cite{Pu19} on data encryption and decryption, and Fig 4(c) illustrates the communication overhead of SOLEK in content transmission is less than that of PCWX \cite{Pan19} and PWXL\cite{Pu19}.}

\begin{figure*} \label{Fig3}
\centering
\subfigure[Time Cost for Server on Encryption]{
\label{Inte1}
\includegraphics[width=0.32\textwidth]{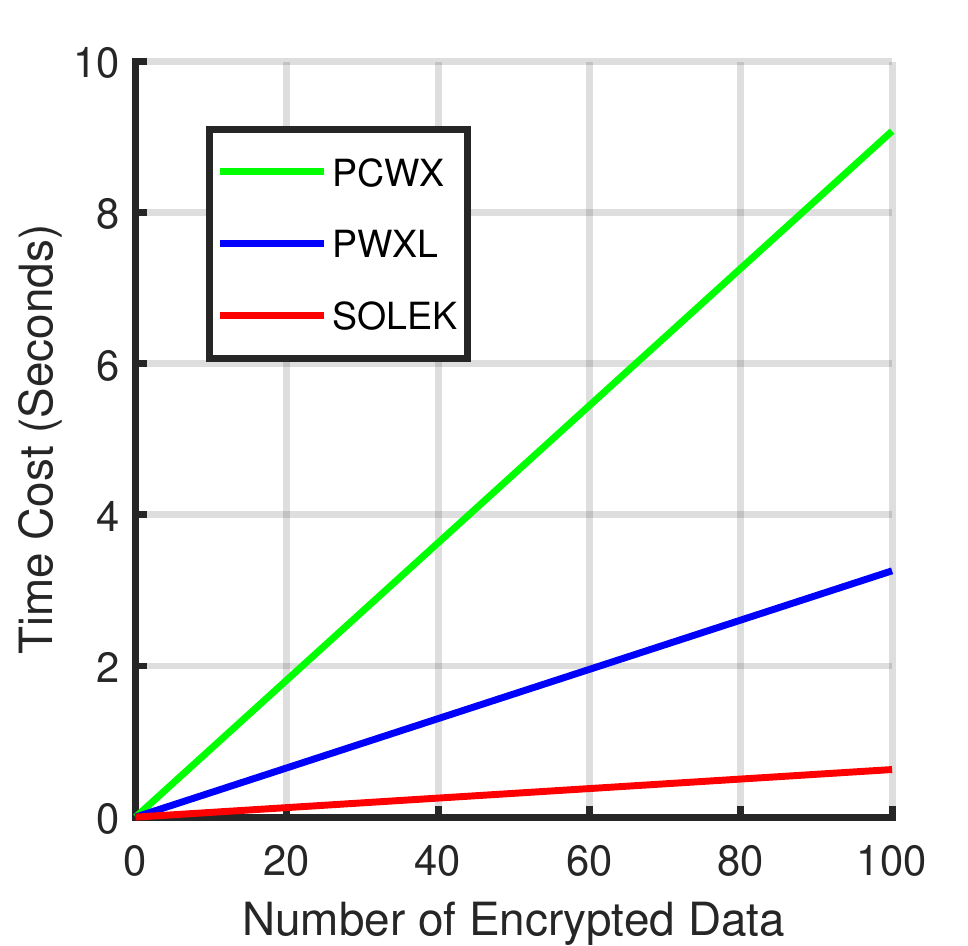}}
\subfigure[Time Cost for UE on Decryption]{
\label{Inte2}
\includegraphics[width=0.32\textwidth]{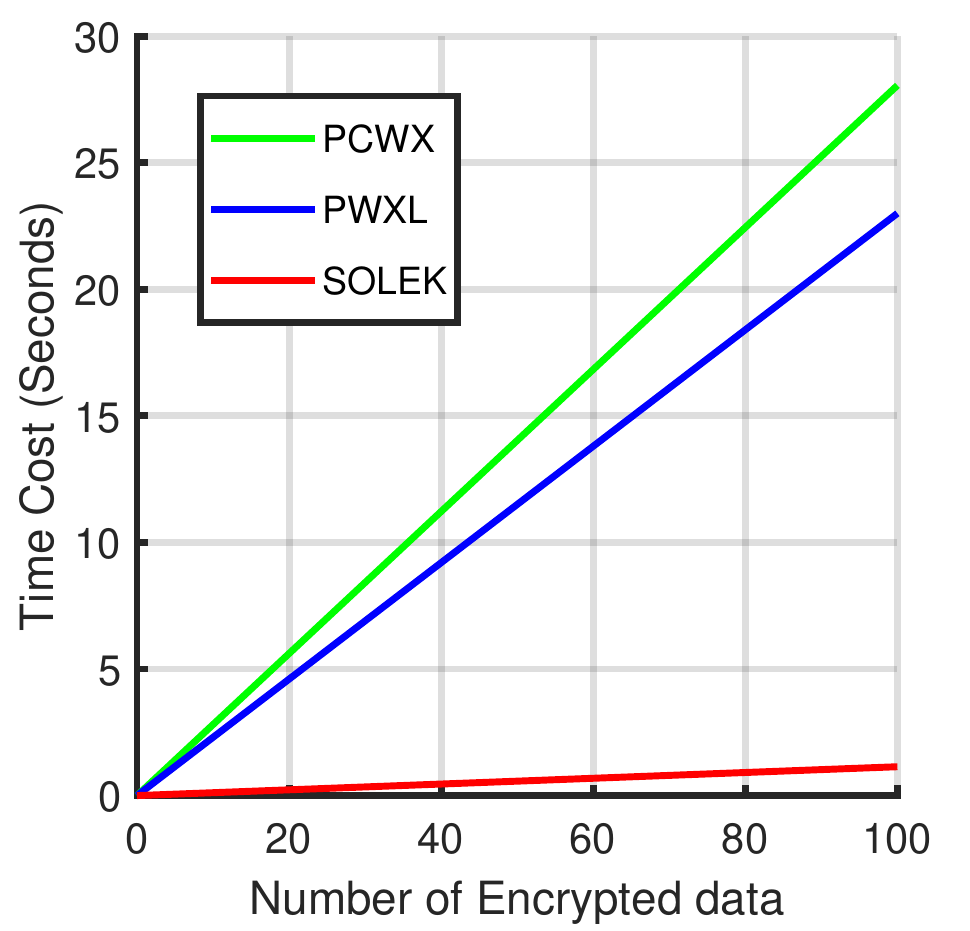}}
\subfigure[Communication Overhead]{
\label{Fig6}
\includegraphics[width=0.32\textwidth]{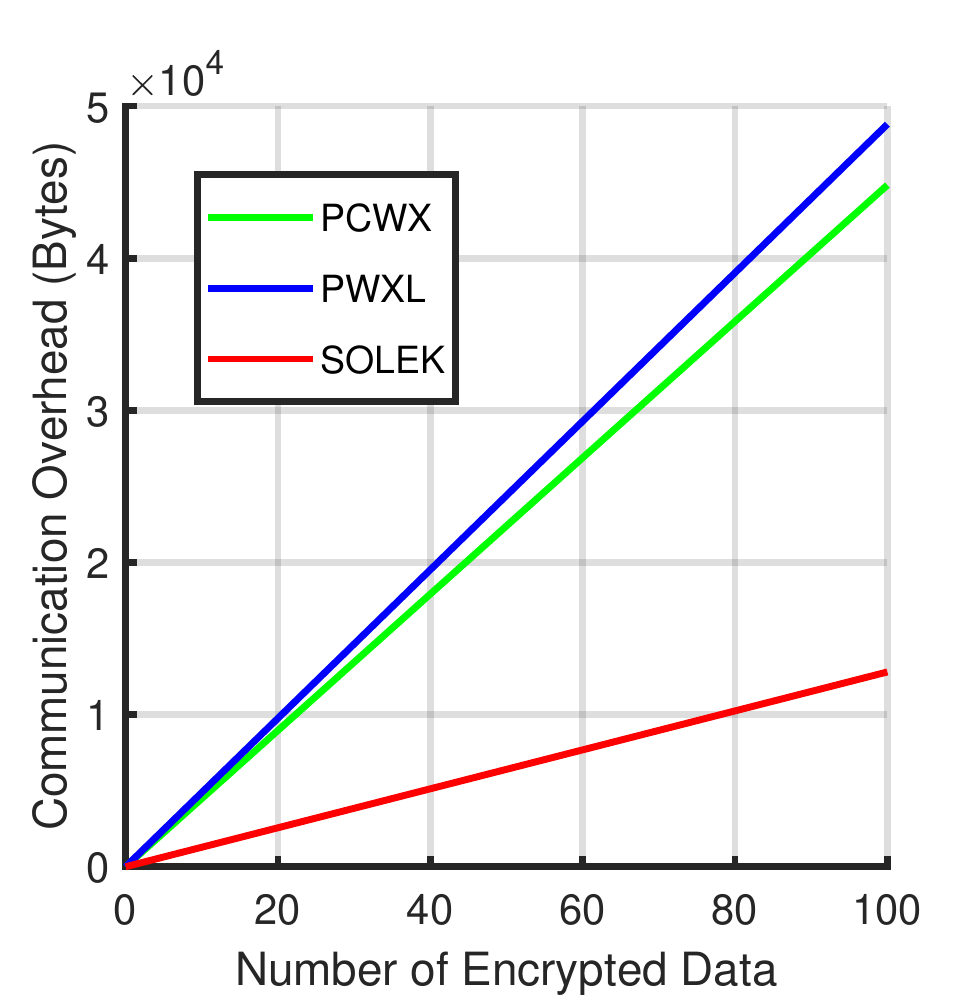}}
\caption{Performance Comparison}
\end{figure*}





\section{Future Directions} \label{sec4}
In this section, we further explore the potential techniques to address the security, privacy, and trust challenges in mobile edge caching, which are expected to attract attentions in the future research.

\subsection{Privacy-preserving Content Placement}
Federated learning \cite{Wang19} can be utilized to predict content popularity. In federated learning, the model parameters are locally trained on edge nodes, and periodically exchanged and updated through unreliable network channels between the edge nodes and the remote server. The private information of mobile users does not need to be sent to the server. However, the information is still vulnerable to being disclosed to edge nodes during historical data collection. {Central differential privacy (DP), multi-party computation, and homomorphic encryption are possible to protect the confidentiality of content, but each has inherent weaknesses. The central DP relies on a trusted ``aggregator" that holds the sensitive data of all individuals for protecting their privacy, multi-party computation requires extra interactions between local edge nodes, and homomorphic encryption causes heavy computational overhead for UE. The privacy leakage may be possible to be prevented by utilizing local differential privacy (LDP), in which individuals randomly perturb their inputs to offer plausible deniability of their data without a trusted party. However, this technique is still in its infancy, although a few of approaches have been proposed to achieve strong privacy preservation in federated learning that support shallow models such as logistic regression. These approaches suffer from the intrinsic limitations that they either cause poor estimation accuracy or only focus on simple datasets or tasks. Moreover, they incur a significant privacy issue of local model, as they expose all internal model states to white-box inference attacks. } The major challenge of LDP is that it introduces large noise to local training in federated learning, such that the cache-hit rate is quite low. Therefore, it is essential to design new data pre-processing mechanisms (e.g., data encoding, and matrix permutation) or integrate distribution estimation techniques for privacy-preserving federated learning in content placement.

\subsection{Efficient and Secure Content Delivery}
The Cuckoo filter is an efficient data structure for membership check to locate the requested contents on edge nodes for mobile users. Compared with the Bloom filter, the Cuckoo filter supports the removal of unpopular content and achieves constant lookup cost for edge nodes. However, false positive matches are still possible. Thus, the naming policies of contents, including flat naming and hierarchical naming, should be extended to avoid false positives. Furthermore, in distributed edge caching, it is challenging to locate the content in the neighbouring edge nodes and determine the time-to-live (TTL) value of a content request. To ensure efficient content localization, an edge node should maintain all Cuckoo filters of the edge nodes within the TTL range that consume the linear storage space of the edge node with the number of reachable edge nodes. To reduce storage costs, the approaches to compressing the elements at the same indices of the Cuckoo filters and encapsulating the content indices for the edge node are deserved to investigate.
In addition, the efficient connection checking mechanisms are required to enable an edge node to check connections with its neighbouring edge nodes periodically and update TTL values.

\subsection{Trustful Content Usage}
The content is cached on an edge node once a mobile user requests, the corresponding signature of the server can be maintained on the public blockchain for authentication and integrity verification. Subsequently, mobile users can check the content integrity based on the signature on the blockchain when needed. Thus, it is unnecessary to forward the signature to the mobile users. Also, secure binding of cache keys and contents (i.e., key-data integrity) should be achieved against potential replace attacks and poisoning attacks. The zero-knowledge proof can be used to ensure that the contents and the cache keys on the blockchain are identical to those cached on the edge nodes. This proof of consistency is essential to ensure content reliability and integrity. Moreover, the behaviors of fog nodes should be regulated. Smart contracts are designed to enable edge nodes to make commitments and enforce punishment on untrusted or rogue edge nodes in case any misbehavior is detected. However, the usage of smart contracts can only guarantee enforcement, not privacy. That is, the privacy of mobile users will be disclosed. Although the variants of zero-knowledge proofs, such as Zero-Knowledge Succinct Non-Interactive Argument of Knowledge (zk-SNARKs) and Zero-Knowledge Succinct Transparent Argument of Knowledge (zk-STARKs) can offer privacy-preserving payments and interactions,
the open problems, including the efficiency and the trusted setup in zk-SNARKs and proof size reduction in zk-STARKs are receiving a lot of attentions in research.

\section{Conclusion} \label{sec6}
{In this article, we have studied the security, privacy, and trust issues in mobile edge caching. In specific, we have answered four key questions in mobile edge caching to have the better understanding, and introduced its security threats towards the content in caches. The challenges of privacy,
security, and trust in content placement, content delivery, and content usage of network edge caching are presented, respectively. In response to efficient and
secure content delivery, a service-oriented and location-based efficient key distribution
protocol (SOLEK) is proposed based on certificateless proxy re-encryption. Finally, several interesting research issues and potential techniques have been identified to shed light on the further research about secure, privacy-preserving, and trustful mobile edge caching.}

\ifCLASSOPTIONcaptionsoff
  \newpage
\fi




\begin{thebibliography}{1}
\bibitem{Stegiou18}
C. Stergiou, K. E. Psannis, B.-G. Kim, and B. Gupta, ``Secure integration of IoT and Cloud Computing," \emph{Future Generation Computer Systems}, vol. 78, no. 3, pp. 964-975, 2018

\bibitem{Hu15}
Y. Hu, M. Patel, D. Sabella, N. Sprecher, and V. Young, ``Mobile Edge Computing -- A Key
Technology Towards 5G," \emph{ETSI White Paper}, vol. 11, no. 11, pp. 1-16, 2015.
\bibitem{Zeydan16}
E. Zeydan, E. Bastug, M. Bennis, M. A. Kader, I. A. Karatepe, A. S. Er, and M. Debbah, ``Big
Data Caching for Networking: Moving From Cloud to Edge," \emph{IEEE Communications Magazine}, vol. 54, no. 9, pp. 36-42, 2016.
\bibitem{Liu16}
D. Liu, B. Chen, C. Yang, and A. F. Molisch, ``Caching at the Wireless Edge: Design Aspects, Challenges, and Future Directions," \emph{IEEE Communications Magazine}, vol. 54, no. 9, pp. 22-28, 2016.

\bibitem{Wang14}
X. Wang, M. Chen, T. Taleb, A. Ksentini, and V. C. Leung, ``Cache in the Air: Exploiting Content Caching and Delivery Techniques for 5G Systems," \emph{IEEE Communications Magazine}, vol. 52, no. 2, pp. 131-139, 2014.


\bibitem{Yang17}
P. Yang, N. Zhang, Y. Bi, L. Yu,  and X. Shen, ``Catalyzing Cloud-Fog Interoperation in 5G Wireless Networks: An SDN Approach," \emph{IEEE Network Magazine}, vol. 31, no. 5, pp. 14--20, 2017.

\bibitem{Lu20}
R. Lu, L. Zhang, J. Ni, and Y. Fang, ``5G Vehicle-to-Everything (V2X) Services: Gearing Up for Security and Privacy," \emph{Proceedings of the IEEE}, vol. 108, no. 2, pp. 373--389, 2020.

\bibitem{Xiao18}
L. Xiao, X. Wan, C. Dai, X. Du, X. Chen, and M. Guizani, ``Security in Mobile Edge Caching with Reinforcement Learning," \emph{IEEE Wireless Communications}, vol. 25, no. 3, pp. 116--122, 2018.

\bibitem{Wang19}
X. Wang, Y. Han, C. Wang, Q. Zhao, X. Chen, and M. Chen, ``In-edge AI: Intelligentizing
Mobile Edge Computing, Caching and Communication by Federated Learning," \emph{IEEE Network
Magazine}, vol. 33, no. 5, pp. 156--165, 2019.



\bibitem{Cohen19}
I. Cohen, G. Einziger, R. Friedman, and G. Scalosub, ``Access Strategies for Network Caching,"
in \emph{Proc. of IEEE INFOCOM}, 2019, pp. 1--9.



\bibitem{Leguay17}
J. Leguay, G. S. Paschos, E. A. Quaglia, and B. Smyth, ``CryptoCache: Network Caching with
Confidentiality," in \emph{Proc. of IEEE ICC}, 2017, pp. 1--6.

\bibitem{Su18}
Z. Su, Q. Xu, J. Luo, H. Pu, Y. Peng, and R. Lu, ``A Secure Content Caching Scheme for Disaster Backup in Fog Computing Enabled Mobile Social Networks," \emph{IEEE Transactions on Industrial Informatics}, vol. 14, no. 10, pp. 4579--4589, 2018.

\bibitem{Pan19}
J. Pan, J. Cui, L. Wei, Y. Xu, and H. Zhong, ``Secure Data Sharing Scheme for VANETs Based
on Edge Computing," \emph{EURASIP Journal on Wireless Communications and Networking}, article no. 169, 2019.


\bibitem{Selvi17}
S. Selvi, D. Sharmila, P. Arinjita, and C. P. Rangan, ``An Efficient Certificateless Proxy Re-encryption Scheme Without Pairing," in \emph{Proc. of ProvSec}, 2017, pp. 413--433.



\bibitem{Pu19}
Y. Pu, Y. Wang, F. Yang, J. Luo, C. Hu, and H. Hu, ``An Efficient and Recoverable Data Sharing Mechanism for Edge Storage," in \emph{Proc. of WASA}, 2019, pp. 247--259.







\end{thebibliography}
%

%

\begin{IEEEbiography}[{\includegraphics[width=1in,height=1.25in,clip,keepaspectratio]{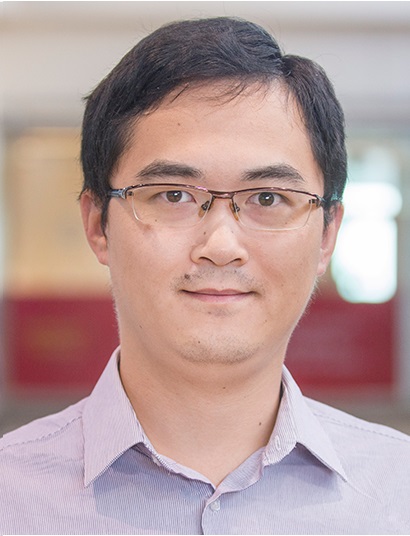}}] {Jianbing Ni}(M'18) is currently an assistant professor with the Department of Electrical and Computer Engineering, Queen's University, Kingston, Canada. He received his Ph.D. degree in Electrical and Computer Engineering from University of Waterloo, Waterloo, Canada, in 2018. His research interests are applied cryptography and network security, with current focus on edge computing, mobile crowdsensing, Internet of Things, and Blockchain technology. \end{IEEEbiography}



\begin{IEEEbiography}[{\includegraphics[width=1in,height=1.25in,clip,keepaspectratio]{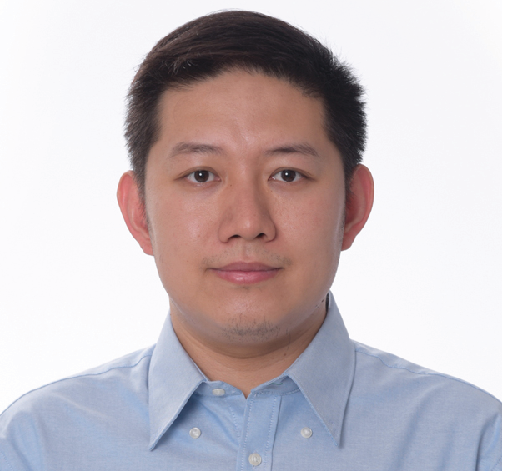}}]{Kuan Zhang} (S'13-M'17) received the Ph.D. degree in Electrical and Computer Engineering from the University of Waterloo, Canada, in 2016. He was also a postdoctoral fellow with Department of Electrical and Computer Engineering, University of Waterloo, Canada. Since 2017, he has been an assistant professor at the Department of Electrical and Computer Engineering, University of Nebraska-Lincoln, USA. His research interests include security and privacy for mobile social networks, e-healthcare systems, cloud computing and cyber physical systems. \end{IEEEbiography}



\begin{IEEEbiography}[{\includegraphics[width=1in, height=1.25in, clip, keepaspectratio]{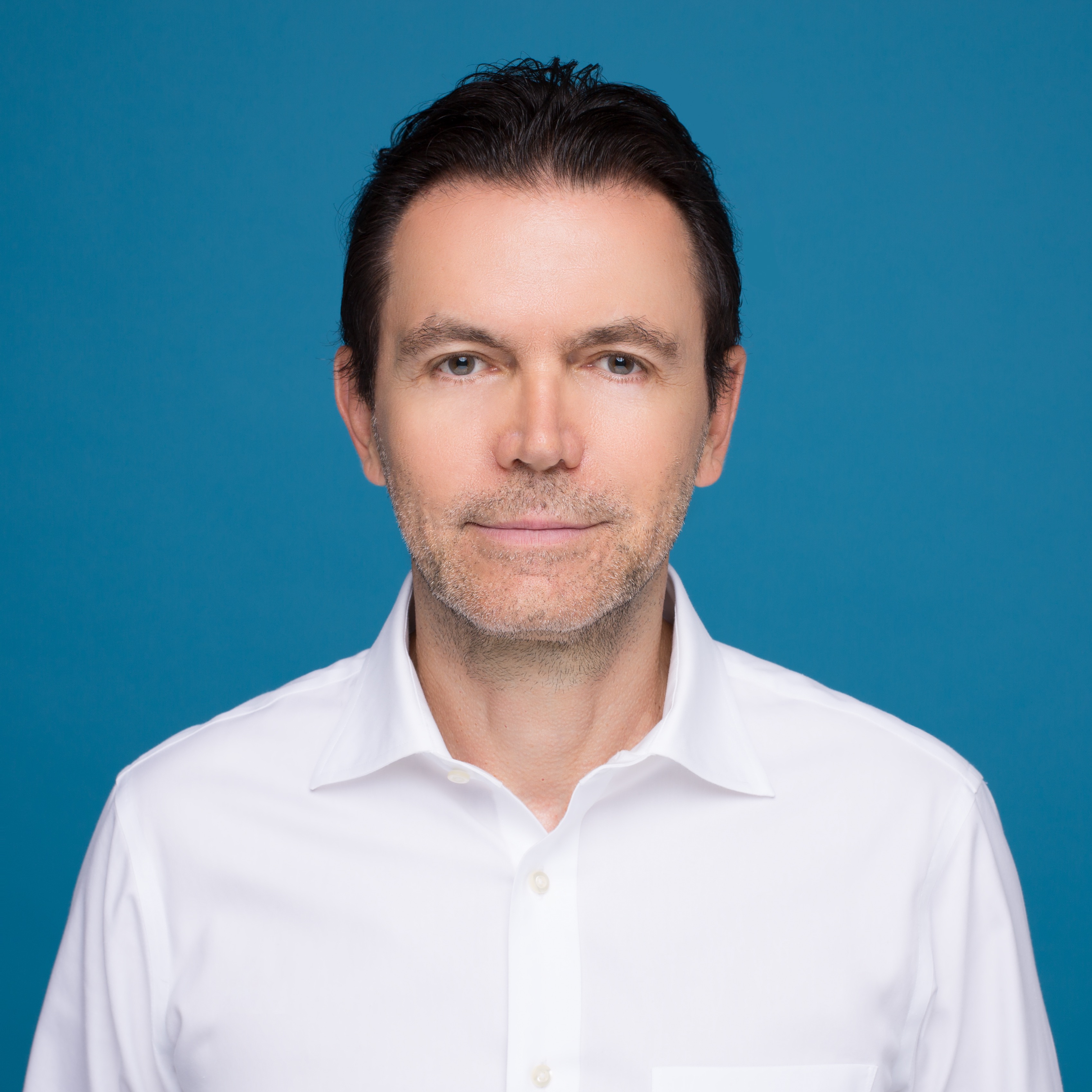}}] {Athanasios V. Vasilakos} is with the School of Electrical and Data Engineering, University Technology Sydney, Australia, and with the Department of Computer Science, Electrical and Space Engineering, Lulea University of Technology, Lulea, 97187, Sweden. He served or is serving as an Editor for many technical journals, such as the IEEE Transactions on Network and Service Management, IEEE Transactions on Cloud Computing, IEEE Transactions on Information Forensics and Security, IEEE Transactions on Cybernetics, IEEE Transactions on NanoBioscience, IEEE Transactions on Information Technology in Biomedicine, ACM Transactions on Autonomous and Adaptive Systems, and IEEE Journal on Selected Areas in Communications. He is Web of Science 2017,2018,2019,2020 Highly Cited Researcher. \end{IEEEbiography}

\end{document}